# Investigation of the thickness-dependence of the Dzyaloshinskii-Moriya interaction in Co$_2$FeAl ultrathin films: effects of the annealing temperature and the heavy metal material

M. Belmeguenai [1,*], Y. Roussigné[1], S. M. Chérif[1] and A. Stashkevich[1]
[1]*LSPM, CNRS-Université Paris 13, Sorbonne Paris Cité, 99 avenue Jean-Baptiste Clément Université Paris 13, 93430 Villetaneuse, France*

M. Nasui[2] and M.Gabor[2,**]
[2]*Center for Superconductivity, Spintronics and Surface Science, Technical University of Cluj-Napoca, Str. Memorandumului No. 28 RO-400114 Cluj-Napoca, ROMANIA*

A. Mora-Hernández[3], B. Nicholson[3], O.-O. Inyang[3] and A.T. Hindmarch[3]
[3]*Department of Physics, Durham University, South Road, Durham DH1 3LE, United Kingdom*

L. Bouchenoire[4]
[4] *XMaS, European Synchrotron Radiation Facility, F-38000 Grenoble, France*

*Abstract-* The interfacial Dzyaloshinskii-Moriya interaction (iDMI) has been investigated in Co$_2$FeAl (CFA) ultrathin films of various thicknesses (0.8 nm$\leq t_{CFA} \leq$2 nm) grown by sputtering on Si substrates, using Pt, W, Ir and MgO buffer or/and capping layers. Vibrating sample magnetometry (VSM) revealed that magnetization at saturation ($M_s$) for the Pt and Ir buffered films is higher than the usual $M_s$ of CFA due to the proximity induced magnetization (PIM) in Ir and Pt, estimated to be 19% and 27%, respectively. The presence of PIM in these materials is confirmed using x-ray resonant magnetic reflectivity. Moreover, while no PIM is induced in W, higher PIM is obtained with Pt when it is used as buffer layer rather than capping layer. Brillouin light scattering (BLS) in the Damon-Eshbach geometry has been used to investigate the thickness dependencies of iDMI constants, from the spin waves non-reciprocity, and the perpendicular anisotropy field versus the annealing temperature. The DMI sign has been found to be negative for Pt/CFA and Ir/CFA while it is positive for W/CFA. The thickness dependence of the effective iDMI constant for stacks involving Pt and W shows the existence of two regimes similarly to that of the perpendicular anisotropy constant due to the degradation of the interfaces as the CFA thickness approaches the nanometer. The surface iDMI and anisotropy constants of each stack have been determined for the thickest samples where a linear thickness dependence of the effective iDMI constant and the effective magnetization has been observed. The interface anisotropy and iDMI constants, investigated for Pt/CFA/MgO system, showed different trends with the annealing temperature. The decrease of iDMi constant with increasing annealing temperature is probably due to the electronic structure changes at the interfaces, while the increase of the interface anisotropy constant is coherent the interface quality and disorder enhancement.

**I- Introduction**

Recent advances in thin film fabrication processes have led to the possibility of the growth of ultrathin multilayers with a high quality interfaces. Consequently, ultrathin systems incorporating heavy metal/ferromagnet (HM/FM) stacks are currently under intensive research due to their potential applications in the field of spintronics. Indeed, various novel mechanisms and phenomena could occur in these structures such as spin Hall [1-4] and inverse spin Hall effects [5-7], spin orbit torques [8] and interface Dzyaloshinskii-Moriya interaction (iDMI) [9, 10]. This latter is an antisymmetric exchange interaction which favours non-collinear alignment of neighbouring spins. The interfacial DMI can be induced by the large spin-orbit coupling of the HM in contact with ultrathin ferromagnetic materials where the inversion symmetry is broken at the surface.

If the iDMI is strong enough, it should change the static, as well as the dynamic, properties of the system. Indeed, it converts the magnetostatically favourable Bloch wall into a chiral Néel wall [11] and induces chiral canting of spins which lead to special chiral spin textures and in particular to magnetic skyrmions [12]. Skyrmions are thus swirling spin textures, which are topologically protected and therefore are considered as potential candidates for future energy efficient spintronic devices [13]. Skyrmions in thin films are induced by the interfacial DMI (if the DM interaction is strong enough relative to other interactions) for which there are several choices of materials. Therefore, it is essential that iDMI is quantified in different material systems, by determining its effective ($D_{eff}$) or surface ($D_s$) constants [14], using reliable techniques to engineer the desired stacks presenting the suitable iDMI constant. Furthermore, Heusler alloys, such as $Co_2FeAl$ having extremely small damping [15], high Curie temperatures and relatively high spin polarization are very attractive materials for spintronics. In such alloys, there are always some degrees of chemical disorder, which strongly influences many of their physical properties [16], especially iDMI. Therefore, an annealing process is required to initiate crystallization and to induce atomic ordering. Since DMI, is sensitive to disorder, defects and atoms arrangement at the interfaces, it is of great interest for applications and fundamental research to investigate the iDMI in Heusler based stacks, where different HM materials are used.

In this paper, we address the thickness dependence of the iDMI in $Co_2FeAl$/HM. Special interest is given to effects of the annealing temperature ($T_a$) and of the HM nature (where the HM used as buffer and/or capping layers are Ir, W and Pt) on iDMI. For this Brillouin light scattering (BLS) technique, where the wave-vector of the spin wave (SW) is determined by the wavelength and the angle of incidence of the laser beam is used. Since the iDM interaction is antisymmetric in its nature, it would lift the degeneracy of SWs propagating along two opposite directions perpendicular to the static magnetization. Therefore, BLS is the most direct and efficient method for DMI characterization since few parameters are required for the experimental data fit and it can detect simultaneously SWs propagating in the opposite directions. We show that Pt and Ir induce iDMI constants with similar sign while it is opposite for W. Moreover, the iDMI constant induced by Pt decreases with the increasing annealing temperature probably due to the electronic structure changes at the interfaces.

**II- Samples and experimental techniques**

$Co_2FeAl$ (CFA) thin films were grown at room temperature on a thermally oxidized Si substrate using a magnetron sputtering system with a base pressure lower than $2\times10^{-8}$ Torr. Prior to the deposition of CFA film, a 2 nm thick Ta seed layer and a HM layer (HM= 3 nm-thick Pt, 3 nm-thick Ir or 4 nm-thick W) were deposited on the substrate. Next, the CFA films, with variable thicknesses (0.8 nm≤$t_{CFA}$≤2 nm), were deposited at room temperature by dc sputtering under an Argon pressure of 1 mTorr, at a rate of 0.1 nm/s. Finally, in order to protect the structure from air exposure, the CFA layers were capped with different materials. Therefore, four sets of samples are considered here: i) Pt-buffered samples capped with MgO(1 nm)/Ta (2 nm), ii) W-buffered samples capped by MgO(1 nm)/Ta (2 nm), iii) W-buffered samples capped with Pt (3 nm) and iv) Ir-buffered samples capped with Ti (2 nm). It should be mentioned that the W layers was grown in highly resistive β phase [17], while the Ir and Pt layers have the expected (111) texture [18]. Moreover, CFA films grown on Si substrates have polycrystalline structure and do not display any in-plane preferential growth direction [19]. In heterostructures investigated here, only the HM layer induces DMI in the CFA ultrathin layers. The measurements presented here were performed at room temperature.

Hysteresis loops under perpendicular and parallel to the films plane applied magnetic fields have been measured for all the samples. Then, the static magnetic parameters including the intrinsic value of the saturation magnetization $M_s$ have been deduced. The in-plane angular dependence of the hysteresis loops did not reveal the presence of the usual fourfold magnetocrystalline anisotropy of CFA [16] confirming the absence of any in-plane preferential growth direction. BLS, in Damon-Eshbach (DE) geometry, where the DMI effect on the SWs non-reciprocity is maximal [20], has been used to determine the iDMI constants for each sample. For this, the magnetic field, sufficiently high to warrant the saturation of the magnetization in the film plane, was applied perpendicular to the incidence plane to measure spectra after counting photons up to 19 hours (especially for the highest incidence angles) for different wave vector values ($k_{SW} = \frac{4\pi}{\lambda}\sin\theta$, where $\theta$ the incidence angle and λ=532 nm the laser wavelength). The Stokes (S) and anti-Stokes (AS) frequencies, detected simultaneously were then determined from Lorentzian fits to the BLS spectra.

X-ray resonant magnetic reflectivity (XRMR) [21] measurements were made on the XMaS beamline at the European Synchrotron Radiation Source. The x-ray energy is tuned to close to the $L_3$ absorption edge of either Pt or Ir, and reversible ~90 % circularly polarised beam is induced using a diamond phase-plate. The reflected intensity $I^{+(-)}$, corresponding to beam helicity parallel (anti-parallel) to the saturating in-plane magnetic field applied in the scattering plane, is measured as a function of the scattering angle $2\theta$. The derived asymmetry, $I^+ - I^- / I^+ + I^-$, is sensitive to proximity induced magnetization (PIM) in the heavy metal layer and is expected to reverse sign when the magnetic field is reversed. The asymmetry is uniformly zero in the absence of PIM.

**III- Results and discussion**

Figure 1 shows the $Co_2FeAl$ thickness dependences of the saturation magnetic moment per unit area for the four sets of samples incorporating the various capping and buffer HM layers. The magnetization at saturation ($M_s$) and the magnetic dead layer thickness ($t_d$) are straightforwardly determined from the linear fits of these data: the slope gives the saturation magnetization, while the horizontal axis intercept gives the extent of the dead layer. The thicknesses of the magnetic dead layer are found to be 0.29 nm, 0.42 nm, 0.51 nm and 0.43 nm for, W/CFA/MgO, W/CFA/Pt, Pt/CFA/MgO and Ir/CFA/Ti, respectively (see tableau 1). The magnetic dead layer is mainly due to intermixing at the HM/CFA interface and a possible oxidation at the CFA/MgO interface. The largest magnetic dead layer thickness has been observed for Pt/CFA/MgO and the smallest one for W/CFA/MgO, suggesting more interdiffusion from the interface Pt/CFA than the W/CFA. $M_s$ values, shown in table 1, change substantially with the stacks. The largest magnetization value ($M_s$=1167±70 emu/cm$^3$) has been observed for Pt/CFA/MgO and the smallest one for W/CFA/MgO ($M_s$=811±50 emu/cm$^3$). For W/CFA/Pt and Ir/CFA/Ti the $M_s$ are 983±50 emu/cm$^3$ and 1063±55 emu/cm$^3$, respectively. While, the $M_s$ of W/CFA/MgO is comparable to that of MgO/CFA/MgO ($M_s$~850±50 emu/cm$^3$) [16], clear enhancement is observed for the other sets of the samples most likely due to the proximity induced magnetization in Pt and in Ir. This corresponds to a change in film magnetization of 13.5%, 19% and 27% for W/CFA/Pt, Ir/CFA/Ti and Pt/CFA/MgO, respectively which are in good agreement with the reported values for Ir/Co and Pt/Co [22] systems. It is worth mentioning that no PIM is induced in W and that the Pt/CFA interface provides higher PIM compared to the CFA/Pt interface. To check the possibility of existence of PIM in the different HM materials, XRMR measurements have been performed. Figure 2 shows XRMR reflected intensity and asymmetry for Ir/CFA(2 nm)/MgO and Pt/CFA(2 nm)/MgO structures. The reflected intensity (upper) has sharper features for the Ir/CFA/MgO structure compared to the Pt/CFA/MgO structure, suggesting slightly sharper structural interfaces in the Ir/CFA/MgO structure. This is in good agreement with the smaller dead layer thickness in Ir/CFA/Ti revealed by VSM. However, the asymmetry signal is generally more pronounced in the Pt/CFA/MgO structure, which indicates a slightly larger PIM in Pt than in Ir adjacent to CFA, consistent with the VSM results in figure 1. The x-ray measurement does not allow us to measure at the W resonance energy. The presence of PIM in the Pt and Ir layers confirms that the measured reduction in magnetization in CFA at the interface with HM is not the result of a magnetically dead layer, but rather a region of reduced magnetization in the vicinity of the interface with HM; a magnetic dead-layer would not produce PIM in the adjacent heavy metal.

Figure 3 shows the typical BLS spectra for the W/CFA (1.2 nm)/MgO, Ir/CFA (1.2 nm)/Ti and Pt/CFA (1.2 nm)/MgO samples for $k_{sw}$=20.45 μm$^{-1}$ ($\theta$=60°) and for different applied fields larger than the saturation field which is sample dependant. Note the lower signal to noise ratio for the W/CFA/MgO, suggesting the lesser quality of CFA grown on W. The spectra reveal the existence of both S and AS spectral lines. A pronounced difference between the frequencies of the S (left line of the spectra) and AS (right line of the spectra) modes ($\Delta F=F_S-F_{AS}$) is revealed by the BLS spectra. This frequency mismatch is HM dependent: for fixed CFA thickness, the larger shift is obtained for Pt/CFA/MgO while W/CFA/MgO provides the smallest value. Moreover, $\Delta F$ is positive for W/CFA/MgO and negative for Ir/CFA/Ti and Pt/CFA/MgO and it is due to the interfacial DMI induced by Pt, Ir and W, as demonstrated previously [14, 23, 24]. Figure 4a shows the $k_{sw}$ dependence of $\Delta F$ for the four sets of the CFA thin films of various thicknesses, where a clear linear behavior can be observed. Note the negative sign of $\Delta F$ for Pt/CFA/MgO and Ir/CFA/Ti and the positive sign for W/CFA/MgO and W/CFA/Pt suggesting that besides the opposite sign of the iDMI induced by W and Pt, the Pt buffer (capping) layer induces a negative (positive) iDMI effective constant. This sign inversion with respect to the stack order confirms the interfacial origin of the DMI and is consistent with the three-sites indirect exchange mechanism [25, 26] proposed previously, confirming thus our previous observations [27]. The $k_{sw}$ dependence of the frequency difference is given by [27]:

$$\Delta F = F_S - F_{AS} = \frac{2\gamma}{\pi M_s} D_{eff} k_{sw} = \frac{2\gamma}{\pi M_s} \frac{D_s}{t_{FM}} k_{sw} \qquad (1)$$

From the slopes of the $k_{sw}$ dependences of $\Delta F$, the effective iDMI constants have been extracted using equation (1) with the gyromagnetic ratio $\gamma/(2\pi)=29.2$ GHz/T (determined previously [15, 19] from ferromagnetic resonance measurements) and the above mentioned values of $M_s$. The variation of $D_{eff}$ versus the reciprocal CFA effective thickness, defined as $1/t_{eff}=1/(t_{CFA}-t_d)$ is shown in figure 4b. Note the deviation from theoretical linearity, as the CFA nominal thickness approaches a critical thickness which is stack dependent: two regimes of a different sign slopes can be distinguished. In the second regime corresponding to thinner CFA films with respect to the critical thickness, $\Delta F$ unexpectedly decreases with the thickness suggesting a degradation of the interfaces for ultrathin films. The linear fit of the data of figure 4b, for CFA thickness where theoretical relation between $D_{eff}$ and $D_s$, given by equation (1) is respected leads to, $D_s$ =-0.33, -0.92, 0.21 and 0.74 pJ/m for Ir/CFA/Ti, Pt/CFA/MgO, W/CFA/MgO and W/CFA/Pt, respectively. These values are significantly lower than that of Pt/Co/AlO$_x$ systems [14] but have the same sign as Pt/Co and confirming the recent results of Kim et al. [23] for Ir/Co. It is worth remembering the positive sign of the iDMI constant for W/CFA/MgO and the good agreement with the obtained value of W/CoFeB/SiO$_2$ ($D_s$=0.21 pJ/m) [28]. Due to the opposite sign of the iDMI constants of W/CFA/MgO and Pt/CFA/MgO, one expects an additive iDMI constants if CFA is sandwiched by W and Pt. However, the weak obtained value of $D_s$ for W/CFA/Pt systems ($D_s$=0.74 pJ/m) leads to iDMI surface constant of 0.53 pJ/m for the CFA/Pt interface. This suggests the lesser quality of this latter interface compared to Pt/CFA one or possibly a different atomic configuration of Pt capping layer.

We have also investigated the thickness dependence of the effective magnetization, defined as: $\mu_0 M_{eff} = \mu_0(M_s - H_k) = \mu_0 M_s - \frac{2K_\perp}{M_s}$, where $H_k$ and $K_\perp$ are the perpendicular uniaxial anisotropy field and anisotropy constant respectively. $M_{eff}$ has been obtained from the fit the mean value of S and AS frequencies using equation (2).

$$F = \left(\frac{F_S + F_{AS}}{2}\right) = \mu_0 \frac{\gamma}{2\pi}\sqrt{(H + Jk_{sw}^2 + P(k_{sw}t_{FM})M_s)(H + Jk_{sw}^2 - P(k_{sw}t_{FM})M_s + M_{eff})} \qquad (2)$$

Where $H$ is the in-plane applied field, $t_{FM}$ is the ferromagnetic layer thickness, $\mu_0$ is the permeability of vacuum and $J = \frac{2A_{ex}}{\mu_0 M_s}$ with $A_{ex}$ is the exchange stiffness constant of CFA.

Figure 5 shows that for all the samples, $M_{eff}$ decreases linearly with the reciprocal effective thickness of CFA, suggesting the existence of perpendicular interface anisotropy. Nevertheless a pronounced non-linear behaviour is observed for the thinner CFA films in W/CFA/MgO stacks. This is in agreement with the poor spectra quality despite a large accumulation time. This feature could be a consequence of the β-W phase for which the normal to the film is not a unique crystallographic direction [17] while buffer Pt and Ir films are (111) oriented [18]. More generally, as mentioned above and due to the interface degradation for thinner CFA films, a clear deviation from the linear behaviour is observed for the stacks showing stronger interface anisotropy. From the slope of the linear fits of the data in figure 5, the interface anisotropy constants, summarized in table 1, have been deduced to be 0.68 mJ/m$^2$, 0.22 mJ/m$^2$, 0.91 mJ/m$^2$ and 0.39 mJ/m$^2$ for W/CFA/MgO, Pt/CFA/W, Pt/CFA/MgO, Ir/CFA/Ti and, respectively. The highest anisotropy constants are obtained for stacks involving MgO. Note also the existence of a non-negligible negative volume perpendicular anisotropy, reinforcing the in-plane magnetization easy axis, estimated to be -0.31 MJ/m$^3$, -0.12 MJ/m$^3$, -0.2 MJ/m$^3$ and -0.1 MJ/m$^3$ for W/CFA/MgO, Pt/CFA/W, Pt/CFA/MgO, Ir/CFA/Ti and, respectively (see Table 1).

The effect of the annealing temperature on both iDMI constants and interface anisotropy has been investigated for the Pt/CFA/MgO stacks showing the higher values for these quantities. The samples have been annealed at 250°C and 400°C in vacuum (with a pressure lower than $3 \times 10^{-8}$ Torr) for 1 hour and then have been characterized by VSM, XRMR and BLS. The VSM measurements showed that the $M_s$ values increase slightly as the annealing temperature increases: $M_s$=1200 ±70 emu/cm$^3$ and 1220 ±70 emu/cm$^3$, respectively for annealing temperature of 250°C and 400°C. Although the $M_s$ variation with respect to that of the as deposited films is within the error bar, we note the slight enhancement of the PIM contribution for annealed samples. It is worth to mention that the magnetic dead layer thickness has decreases to 0.38 nm for the annealed films at 400°C. We thus conclude on the increase of the proximity magnetization in Pt with the enhancement of the interface quality.

From the XRMR measurements of the Pt/CFA(2 nm)/MgO sample annealed at 400°C and shown in figure 6, we see changes in both the reflected intensity and asymmetry after annealing, indicative of both structural and magnetic changes in the structure. At lower scattering angles, below 2 degrees, the structure is largely unchanged after annealing at 400°C but the asymmetry signal is significantly enhanced, indicative of an increased PIM. Note that the PIM itself is small and localised close to the interface [29], such that this enhancement makes a low contribution to the $M_s$ value measured by VSM (that is less adapted to evidence PIM). At scattering angles from around 3 degrees upwards we see changes in the reflected intensity which suggest subtle changes in the multilayer structure. The more pronounced Keissig fringes in the annealed structure suggests a sharper interface between Pt and CFA after annealing at 400°C which is in agreement with the smaller dead layer thickness deduced from VSM. We thus conclude on the increase of the proximity magnetization in Pt with the enhancement of the interface quality. This could be related to the reorganisation of the Fe and Co atoms yielding an enhancement of the Pt atoms polarisation as well as a structural improvement of the interfaces.

The typical BLS spectra measured for $k_{sw}$=20.45 µm$^{-1}$ are shown in figure 7 for the 1.2 nm thick films and for the three different annealing temperatures. One can observe a significant enhancement of the signal to noise ratio as the annealing temperature increases. A significantly decrease of the full width at half maximum linewidth ($\delta F$) as the annealing temperature increases has been observed: for example for the 1 nm thick film, the mean value of $\delta F$ decreases from 5.8 GHz for the as deposited film to 4.5 GHz for the one annealed at 400°C. This is in agreement with the previously obtained results where we reported a decrease of the damping in CFA films as the annealing temperature increases [19]. Figure 8 shows the thickness dependencies of $D_{eff}$ and $M_{eff}$ of the as deposited and the annealed samples. Films annealed at 400°C show a significant difference with the as deposited and the 250°C annealed samples. Indeed, $D_{eff}$ shows a noticeably decrease at $T_a$ =400°C (figure 8a), where the significant changes in the structural, chemical order and in the magnetic properties start to occur [16]. Furthermore, figure 8 reveals that $D_{eff}$ and $M_{eff}$ follow a linear variation which is annealing temperature dependent. Interestingly, this thickness dependence deviates from the previous linear behaviour for the ultrathin (approaching 1 nm) CFA films and the deviation becomes more pronounced for $T_a$=400°C. This non-linear behaviour is most probably due to the degradation of the interface for ultrathin films as the samples are annealed at higher temperatures. From the linear fit of the experimental data of figure 8 the surface iDMI and the interface anisotropy constants have been deduced. We find that the anisotropy and iDMi constants follow an opposite trends. While the interface anisotropy increases slightly with $T_a$ (1.14 and 1.25 mJ/m$^2$ for $T_a$=250°C and 400°C, respectively), $D_s$ shows drastic decreases for the films annealed at 400°C ($D_s$=-0.85 and -0.61 mJ/m$^2$ for $T_a$=250°C and 400°C, respectively). The different parameters obtained from BLS and VSM data are summarized in table 1. Note the similar trend of PIM and surface anisotropy with the annealing temperature. The increase of anisotropy constant is coherent with the previous observed trends in MgO/CFA/MgO [16]. It is probably due to the enhancement of the CFA/MgO surface anisotropy. While the surface anisotropy results from the two interfaces, the iDMI effect is only due to the Pt/CFA interface. iDMI is sensitive to the Co, Fe electronic structure at the interface which is highly temperature dependent. On the other hand, it is not surprising to observe different trends for iDMI constant and surface anisotropy because the first involves only one interface while the second involves both interfaces. The different behaviour of PIM and iDMI versus annealing temperature probably suggests a different origin in these samples.

**Conclusions**

$Co_2FeAl$ films of various thicknesses were prepared by sputtering on $Si/SiO_2$ substrates using a different buffer and capping layers (MgO, W, Pt and Ir). The vibrating sample magnetometry measurements revealed that in contrast to W, Pt and Ir buffered layers show higher magnetization at saturation, possibly due to the proximity magnetization which we have shown, using x-ray resonant magnetic reflectivity, to be induced in Pt and Ir. We also showed that the Pt buffer layer induces higher proximity magnetization than the Pt capping layer. Brillouin light scattering has been used in the Damon-Eshbach geometry to investigate the spin waves non reciprocity induced by the interfacial Dzyaloshinskii-Moriya interaction (iDMI). It turned out that the iDMI effective constant sign and strength is material and stack order dependent. While Pt and Ir present the same iDMI constant sign, W induces iDMI of opposite sign. Moreover, the Pt layer providing the higher iDMI constant with respect to Ir and W seems to induce higher iDMI effect when it is used as buffer layer rather than capping layer. The effect of the annealing temperature on the iDMI has been studied in $Pt/Co_2FeAl/MgO$ stacks, where a decrease of the iDMI constant has been observed. The analysis of BLS measurements revealed the existence of perpendicular interface anisotropy which shows a different trend with the annealing temperature from that of the iDMI constant.

**Acknowledgements**

We would like to thank EPSRC for the provision of XMaS beamtime. This work has been supported by the Conseil regional d'Île-de-France through the DIM NanoK (BIDUL project). M.S.G and T.P. acknowledge the financial support of UEFISCDI through PN-II-RU-TE-2014-1820 — SPINCOD research Grant No. 255/01.10.2015. AM-H acknowledges fellowship support from CONACYT, BN acknowledges support via an EPSRC DTP award, and O-OI acknowledges support from a Nigerian government TETFund scholarship.

| Stack | W/CFA/MgO | W/CFA/Pt | Pt/CFA/MgO | | | Ir/CFA/Ti |
|---|---|---|---|---|---|---|
| $T_a$ (°) | RT | RT | RT | 250°C | 400°C | RT |
| $t_d$ (nm) | 0.29 | 0.42 | 0.51 | 0.43 | 0.38 | 0.43 |
| $M_s$ (emu/cm$^3$) | 811 | 983 | 1167 | 1200 | 1220 | 1063 |
| $D_s$ (pJ/m) | 0.21 | 0.74 | -0.92 | -0.85 | -0.61 | -0.33 |
| $K_v$ (MJ/m$^3$) | -0.31 | -0.12 | -0.2 | -0.55 | -0.27 | -0.1 |
| $K_s$ (mJ/m$^2$) | 0.684 | 0.224 | 0.91 | 1.14 | 1.25 | 0.39 |

Table I: Parameters obtained from the best fits of the thickness dependencies of the magnetic moment per area unit, the effective magnetization and the effective iDMI constant of the CFA thin films grown on Si substrates using various capping and buffer layers and annealed at different temperatures.

**Fig. 1 : Belmeguenai et al.**

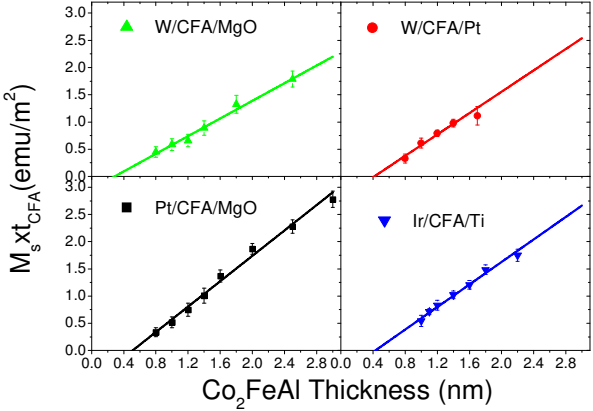

**Fig. 2 : Belmeguenai et al.**

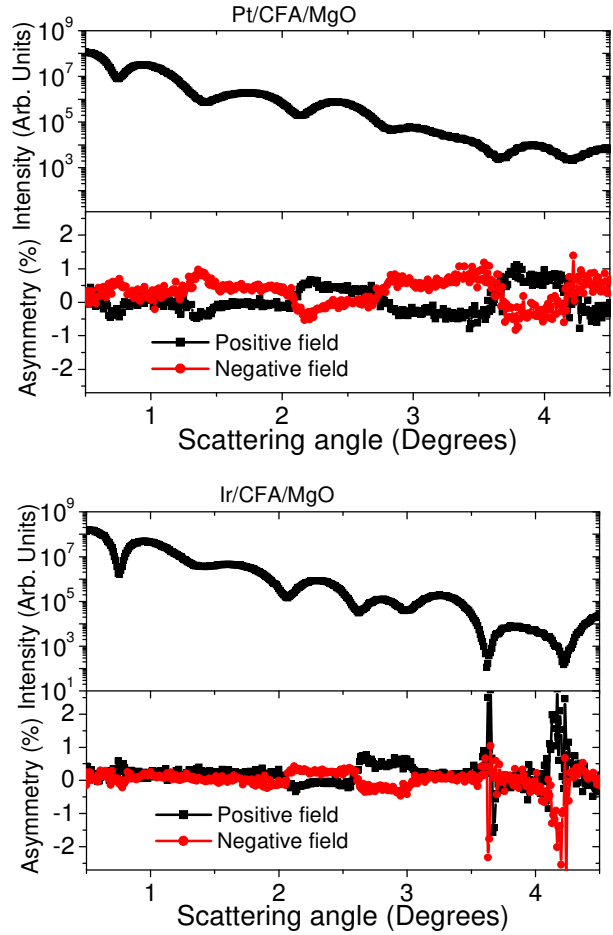

**Fig. 3 : Belmeguenai et al.**

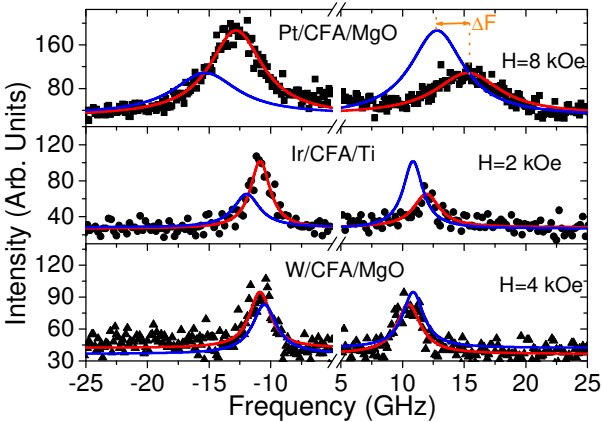

**Fig. 4 : Belmeguenai et al.**

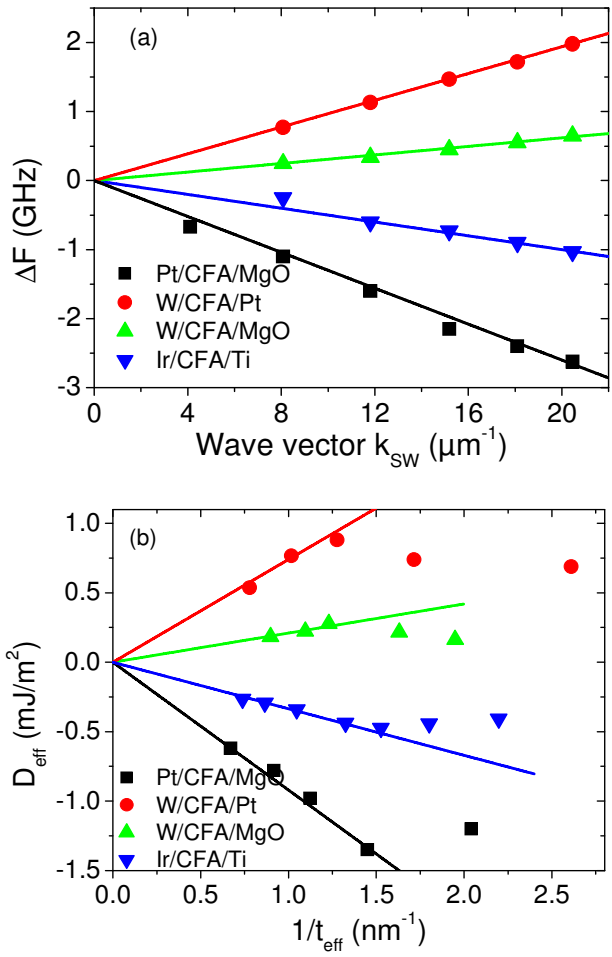

**Fig. 5 : Belmeguenai et al.**

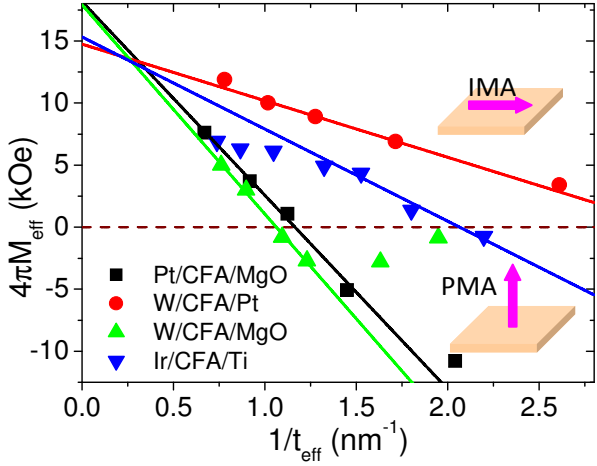

**Fig. 6 : Belmeguenai et al.**

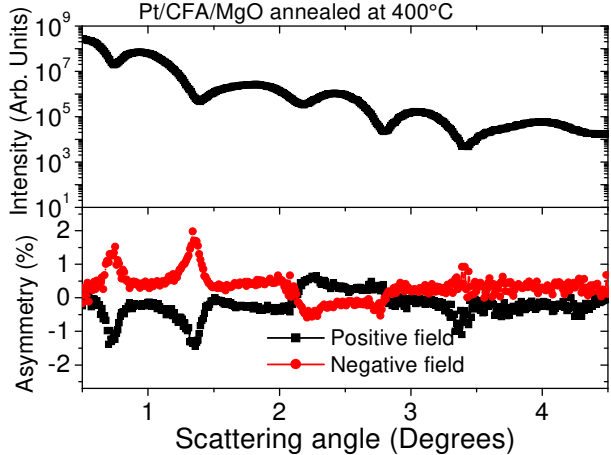

**Fig. 7 : Belmeguenai et al.**

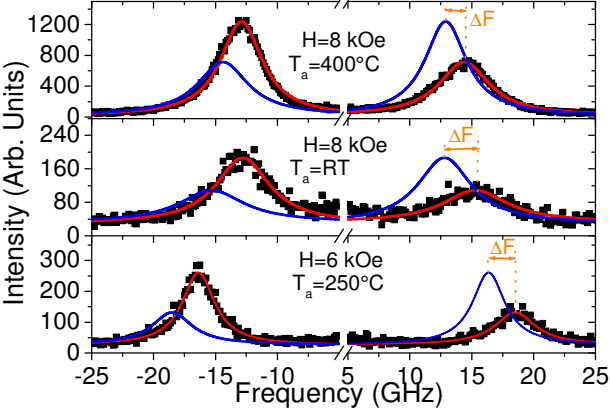

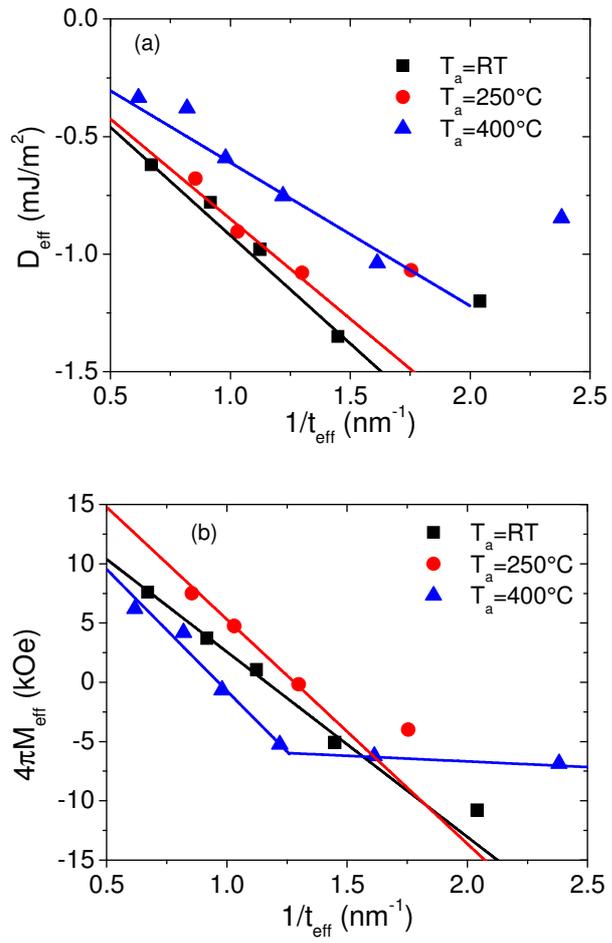

**Fig. 8 : Belmeguenai et al.**

**Figure captions**

**Figure 1:** (Color online) Saturation magnetic moment per unit area versus of $Co_2FeAl$ thickness. $Co_2FeAl$ thin films were grown on Si substrates with various buffer and capping materials. Symbols refer to the VSM measurements and solid lines are the linear fits.

**Figure 2:** (Color online) Specular reflectivity and spin asymmetry data obtained from the x-ray resonant magnetic reflectivity measurements of Pt/CFA(2 nm)/MgO and Ir/CFA(2 nm)/MgO.

**Figure 3:** (Color online) BLS spectra measured for Pt/CoFe(1.2 nm)/MgO, W/CoFe(1.2 nm)/MgO and W/CoFe(1.2 nm)/ Pt films, grown on Si, at 8 kOe, 4 kOe and 2 kOe in-plane applied field, respectively and for a characteristic light incidence angle corresponding to $k_{sw}$ = 20.45 $\mu m^{-1}$. Symbols refer to the experimental data and solid lines are the Lorentzian fits. Fits corresponding to negative applied fields (blue lines) are presented for clarity and direct comparison of the Stokes and anti-Stokes frequencies.

**Figure 4:** (Color online) (a) Wave vector ($k_{sw}$) dependence of the experimental frequency difference $\Delta F$ of Pt/$Co_2FeAl$(1.2 nm)/MgO, W/$Co_2FeAl$(1.2 nm)/MgO, W/$Co_2FeAl$(1.2 nm)/ Pt and Ir/$Co_2FeAl$(1.2 nm)/Ti stacks grown on Si substrates. Symbols are experimental data and solid lines refer to linear fit using equation (1) and magnetic parameters in the text. (b) Variation of the effective iDMI constant ($D_{eff}$) versus the reciprocal effective thickness of $Co_2FeAl$ films in the four sets of stacks with various buffer and capping layers. Symbols refer to measurements and solid lines are the linear fits. $D_{eff}$ values have been obtained from fits of $k_{sw}$ dependence of $\Delta F$ equation (1) and magnetic parameter in the text.

**Figure 5:** (Color online) Effective magnetization ($4\pi M_{eff}$) versus the effective thickness of $Co_2FeAl$ films in Pt/$Co_2FeAl$/MgO, W/$Co_2FeAl$/MgO, W/$Co_2FeAl$/Pt and Pt/$Co_2FeAl$/Ir grown on Si substrates. $4\pi M_{eff}$ values have been extracted from the fit of BLS measurements of the mean

frequency of Stokes and anti-Stokes lines versus the wave vector using equation (2) and parameters in text. Symbols refer to experimental data while solid lines are the linear fits. The horizontal dashed line at zero is used to indicate the thickness region where samples are perpendicularly magnetized ($4\pi M_{eff}$ negative) and in-plane magnetized (positive $4\pi M_{eff}$). The sketch in the inset refers to thicknesses where samples present perpendicular magnetic anisotropy (PMA) and in-plane magnetic anisotropy (IMA).

**Figure 6:** (Color online) Specular reflectivity and spin asymmetry data obtained from the x-ray resonant magnetic reflectivity measurements of Pt/CFA(2 nm)/MgO annealed at 400°C.

**Figure 7:** (Color online) BLS spectra of Pt/CoFe(1.2 nm)/MgO grown on Si and annealed at various temperatures ($T_a$) measured at 8 kOe and 6 kOe in-plane applied field and for a characteristic light incidence angle corresponding to $k_{sw}$ = 20.45 $\mu$m$^{-1}$. Symbols refer to the experimental data and solid lines are the Lorentzian fits. Fits corresponding to negative applied fields (blue lines) are presented for clarity and direct comparison of the Stokes and anti-Stokes frequencies.

**Figure 8:** (Color online) Variation of the (a) effective iDMI constant ($D_{eff}$) and (b) effective magnetization ($4\pi M_{eff}$) versus the reciprocal effective thickness of Co$_2$FeAl films in Pt/CoFe/MgO stacks annealed at various temperatures ($T_a$). Symbols refer to measurements and solid lines are the linear fits. $D_{eff}$ values have been obtained from fits of $k_{sw}$ dependence of $\Delta F$ using equation (1) and magnetic parameter in the text while $4\pi M_{eff}$ values have been obtained from the fit of BLS measurements of the mean frequency of Stokes and anti-Stokes lines versus the wave vector using equation (2).